\newcommand{\kms}{\mbox{km~s$^{-1}$}}

\documentclass{nature}

\usepackage{multibib}
\usepackage{graphicx}
\newcites{art,met}{References, Additional References}

\bibliographystyleart{naturemag}
\bibliographystylemet{naturemag}

\title{The double-degenerate, super-Chandrasekhar nucleus of the planetary nebula Henize~2--428}

\author{M. Santander-Garc\'\i a$^{1,2}$, P. Rodr\'iguez-Gil$^{3,4}$, R. L. M. Corradi$^{3,4}$,  D. Jones$^{3,4}$, B. Miszalski$^{5,6}$, H. M. J. Boffin$^{7}$, M. M. Rubio-D\'\i ez$^{8}$   \&  M. M. Kotze$^{5}$}

\begin{document}
\maketitle

\begin{affiliations}
 \item Observatorio Astron\'omico Nacional, Ap.\ de Correos 112, E-28803, Alcal\'a de Henares, Spain
 \item Instituto de Ciencia de Materiales de Madrid (CSIC), E-28049, Madrid, Spain
 \item Instituto de Astrof\'\i sica de Canarias, E-38200 La Laguna, Tenerife, Spain
 \item Departamento de Astrof\'\i sica, Universidad de La Laguna, E-38205 La Laguna, Tenerife, Spain
 \item South African Astronomical Observatory, PO Box 9, Observatory, 7935, South Africa
 \item Southern African Large Telescope Foundation, PO Box 9, Observatory, 7935, South Africa
 \item European Southern Observatory, Alonso de C\'ordova 3107, 19001 Casilla, Santiago, Chile
 \item Centro de Astrobiolog\'\i a, CSIC-INTA, E-28850 Torrej\'on de Ardoz, Spain
\end{affiliations}


\begin{abstract}
The planetary nebula (PN) stage is the ultimate fate of stars with mass 1 to 8 solar masses (M$_\odot$). The origin of their complex morphologies is poorly understood\citeart{balick02}, although several mechanisms involving binary interaction have been proposed\citeart{demarco09}$^,$\citeart{garciaarredondo04}. In close binary systems, the orbital separation is short enough for the primary star to overfill its Roche lobe as it expands during the Asymptotic Giant Branch (AGB) phase. The excess material ends up forming a common-envelope (CE) surrounding both stars. Drag forces would then result in the envelope being ejected into a bipolar PN whose equator is coincident with the orbital plane of the system. Systems in which both stars have ejected their envelopes and evolve towards the white dwarf (WD) stage are called double-degenerates. Here we report that Henize~2--428 has a double-degenerate core with a combined mass unambiguously above the Chandrasekhar limit of 1.4 M$_\odot$. According to its short orbital period (4.2 hours) and total mass (1.76 M$_\odot$), the system should merge in 700 million years, triggering a Type Ia supernova (SN Ia) event. This finding supports the double-degenerate, super-Chandrasekhar evolutionary channel for the formation of SNe Ia\citeart{howell06}.
\end{abstract}

The hypothesis of binarity as a key to produce bipolar PNe has been progressively gaining ground, particularly with the recent discovery of many new close binary PN nuclei\citeart{boffin11}$^,$\citeart{boffin12}$^,$\citeart{jones14}$^,$\citeart{corradi14}. To this end, we have undertaken a campaign of photometric monitoring of PN central stars that show characteristic features such as rings and/or jets highlighted as being associated with central star binarity\citeart{miszalski09b}. One of them, Henize~2--428 (PN G049.4+02.4), shows one of the shortest orbital periods found for this class of object. This bipolar nebula consists of two open lobes emerging from a ring-shaped waist inclined $\sim$68$^\mathrm{o}$ to the plane of the sky (as deduced from the elliptical shape of the ring in images of the nebula, see Fig.~1). The kinematic age of the bipolar lobes and ring implies that Henize~2-428 is an evolved PN. However, the presence of a high density nebular core indicates that strong mass loss and/or mass transfer is occurring close to its central star, which in turn suggests that the nebula may host a binary core\citeart{rodriguez01}.


The periodic light curves of Henize~2--428 in the Sloan i and Johnson B bands (respective effective wavelengths 0.78 and 0.44 $\mu$m) demonstrate its binary nature (Fig. 2). They show two broad minima indicative of ellipsoidal modulation due to tidal distortion of one or both components. The period analysis using Schwarzenberg-Czerny's\citeart{Scwarzenberg96}  analysis-of-variance (AOV) method on the photometric data set with the greatest time coverage reveals that the orbital period is 0.1758$\pm$0.0005 day. Spectroscopic observations with FORS2 on the Very Large Telescope (Fig. 3) show the absorption lines to be double and varying with the orbital phase. The sinusoidal Doppler shifts of the double He  {\sc ii} 541.2~nm absorption feature are finally revealed by a series of GTC/OSIRIS spectra obtained along a full orbit (Fig. 4). Gaussian fitting of the absorption-line profiles followed by sinusoidal fitting to the data folded on the orbital period indicates that both radial velocity amplitudes are identical, with values of 206$\pm$8 and 206$\pm$12 \kms, respectively. The very similar depth of the each light curve minima, of the intensity of the He {\sc ii} 541.2~nm stellar absorption line, and of the velocity amplitudes indicate that the two stars have nearly identical masses and effective temperatures. The latter can be constrained by the presence of the photoionised nebula and by the lack of He II emission lines in the spectrum of the nebula. The effective temperatures of the stars have therefore been kept between 20,000 and 40,000 K in our modelling. 

The combined analysis of the light curves and radial velocity curves\citeart{wilson79}$^,$\citeart{prsa11} yields the orbital parameters of the system (Table 1). The result is a double-degenerate, overcontact binary, in which both stars overfill their Roche lobes. The total luminosity of the system, 845 L$_\odot$, is compatible with the 690 L$_\odot$ reported in the literature\citeart{stanghellini02}.  

The inclination of the orbital plane is confined to a narrow range between 63.4$^\mathrm{o}$ and 66.1$^\mathrm{o}$ (see table 1), close to the $\sim$68$^\mathrm{o}$-inclined equatorial ring of the nebula, further strengthening the binary hypothesis. It is noteworthy that the nebular and orbital axes of {\it every} PN in which both are known happen to be coincident within a few degrees. These are A41, A63, A65, HaTr4, NGC~6337, NGC~6778, and Sp1\citeart{jones14b}. This is convincing evidence for the binary scenario formation of PNe outlined above.

The most striking result is the total mass of the system, which amounts to 1.76 M$_\odot$. This is the only reported double-degenerate system whose total mass is above the Chandrasekhar limit without any ambiguity\citeart{tovmassian10}, even when considering the lowest mass allowed within the confidence range (1.5 M$_\odot$). Moreover, with a mass ratio (mass of the secondary over mass of the primary) of 1, both stars are quite massive compared to average post-AGB stars, and their luminosity and effective temperatures do not match the theoretical post-AGB cooling tracks\citeart{bloecker95}. The explanation must be sought in their binary evolution. The system seems to have first undergone a phase of mass transfer via wind or stable Roche lobe overflow (RLOF), and then a CE. This is likely, as in order to have two oversized pre-WD stars with R=0.68-0.7 R$_\odot$ still hot, the two events must have happened fast and consecutively. In addition, for the system to have a current mass ratio close to 1, the initial mass ratio must have been also very close to 1. Such a case would lead to a stable RLOF initially, given the relatively massive cores and judging by stability studies\citeart{chen08}. After the first pre-WD star was formed (from a quite massive AGB\citeart{weidemann00}), the mass ratio (AGB/pre-WD) would then be much larger than 1, as the secondary is still a massive AGB as well, and the system would undergo a CE phase. Having an initial mass ratio close to 1 would also explain the final, similar luminosities and effective temperatures. 

An alternate model in which one or both stars are non-degenerate helium stars deserves further discussion. A 0.88 M$_\odot$ helium star approaching core helium exhaustion\citeart{paczynski71}, produced by the hydrogen shell of a massive star being stripped off during RLOF, would have a compatible luminosity of 422 L$_\odot$, although its radius, 0.16 R$_\odot$ is still lower than the 0.68 R$_\odot$ radius found in this work, and the position of the stars of Henize~2--428 in log $g$--$T$ diagrams does not match the locus of helium stars in binary systems\citeart{stroeer07} in the same way as it does not match the aforementioned post-AGB cooling tracks. This alternate model is, however, very unlikely, considering that {\it (a)} the computed He/H and the supersolar N/O abundance values\citeart{rodriguez01} indicate a composition more typical of a processed, AGB envelope than a RGB envelope; and {\it (b)} gas densities in the extended equatorial ring are around 10$^3$ cm$^{-3}$, implying recombination times of the order of a few tens of years for H, and a few years for  N+, O+, and O++. Given the nebula age, of the order of a few thousand years\citeart{rodriguez01}, that excludes the possibility that the nebula has a fossil photoionization. In any case, more robust theoretical models should help understanding the peculiarities of this intriguing system.



The stars are massive and close enough to merge in a Hubble time\citeart{shapiro83} ($\sim$700 million years), thus making Henize~2--428 a strong SN Ia candidate. Such a massive double-degenerate system may well explain supernova remnants like SNR 0509-67.5, in which the single-degenerate scenario can be ruled out\citeart{schaefer12}, and confirms the plausibility of double-degenerate scenarios as SN Ia progenitors.

The thermal time scale of the stars is $\sim$80,000 years, much less than the merging time of the system. If the radius of each star contracts on the thermal timescale, then the system will relatively quickly become detached and the orbital period will shrink accordingly. If this is the case, long-term monitoring of this system might reveal small changes in the orbital period over the course of the next decades.

In addition to Henize~2--428, only half a dozen double-degenerate binary systems are known so far: V458~Vul\citeart{rodriguez10}; SBS~1150+599A, the nucleus of TS~01\citeart{tovmassian10}; MT~Ser, the nucleus of Abell 41\citeart{bruch01}, NGC~6026\citeart{hillwig10}, and Fg~1\citeart{boffin12}. However, estimates indicate they could amount up to 25\% of the total PNe hosting a close-binary central star\citeart{hillwig13}. This figure could be even larger given the strong selection effects against double degenerates in photometric surveys ---in fact, most double-degenerate systems may only be identified via radial velocity studies\citeart{boffin12}. It is clear that a larger sample of double-degenerate central stars of PNe would not only help better assess their contribution to the total number of PNe or SN Ia progenitors, but also better understand the role of common-envelope and/or contact-binarity in the evolution and shaping of PNe and in the production of SN Ia events.



\bibliographyart{msantander}



\begin{addendum}
 \item This work is based on
observations made with the 1~m SAAO, 1.2~m MERCATOR, 2.5~m INT, 4.2~m WHT, 8.2~,m VLT and 10.4~m GTC telescopes. We are grateful to Tom Marsh for the use of PAMELA and MOLLY, to Todd Hillwig, Onno Pols and Javier Alcolea for their valuable comments and to Jorge Garc\'\i a-Rojas and Cristina Zurita for the INT/WFC service observations. This work was partially supported by the spanish MINECO within the grants CSD2009--00038, AYA2012--35330, RYC--2010--05762 and AYA2012--38700.
 \item[Author contributions] MSG, PRG, DJ, MMRD, HB and MMK conducted the observations at the various telescopes. MSG, PRG, DJ and MMK reduced the data. MSG performed the light and radial velocity curves modelling, and wrote the paper. All authors discussed the results and implications and commented on the manuscript at all stages.
 \item[Author information] Reprints and permissions information is available at www.nature.com/reprints. The authors declare that they have no competing financial interests. Correspondence and requests for materials
should be addressed to m.santander@oan.es.
\end{addendum}



\begin{table}
\begin{center}
\footnotesize
{\centering
\caption{Orbital and physical parameters of the nucleus of Henize~2--428}
\medskip
\begin{tabular}{lc}
\hline
Parameter & Model  \\
\hline
Orbital period, $P$ (day)                                           & 0.1758$\pm$0.0005 \\
Epoch of minimum light in the i band (HJD)          &    2456507.53797$\pm$0.00034 \\
Epoch of minimum light in the B band  (HJD)       &     2456484.60636$\pm$0.00077 \\
Epoch of zero phase in radial velocity data (HJD)          &     2456516.50377$\pm$0.026 \\
Mass ratio, $q\equiv\ $M$_2$/M$1$               &    1      \\ 
Orbital separation, $a$ (R$_\odot$)               &    1.59 $\pm$ 0.06  \\   
Inclination, $i$ ($^\mathrm{o}$)                   &    64.7 $\pm$ 1.4      \\ 
Centre-of-mass redshift, $\gamma$ (\kms) &    71 $\pm$ 7            \\ 
Primary mass, $M_1$ (M$_\odot$)             &     0.88 $\pm$ 0.13   \\ 
Secondary mass, $M_2$ (M$_\odot$)             &     0.88 $\pm$ 0.13   \\ 
Primary radius, $R_1$ (Pole, R$_\odot$)    &    0.68 $\pm$ 0.04   \\ 
Secondary radius, $R_2$ (Pole, R$_\odot$)    &    0.68 $\pm$ 0.04   \\ 
Primary Surface Potential\citeart{wilson79}, $\Omega_1$      &    3.50 $\pm$ 0.06   \\ 
Secondary Surface Potential\citeart{wilson79}, $\Omega_2$      &    3.50 $\pm$ 0.06   \\ 
Primary Temperature, $T_1$  ($\times$10$^3$ K)   &   32.4 $\pm$ 5.2  \\ 
Secondary Temperature, $T_2$  ($\times$10$^3$ K)   &   30.9 $\pm$ 5.2  \\
Distance to the system, $d$ (kpc)                                        & 1.4$\pm$0.4 \\
\hline
\end{tabular}
}
\end{center}
\vskip5pt
{\small Orbital and physical parameters of the nucleus of Henize~2--428. These have been determined by modelling the INT/WFC Sloan i and the SAAO 1m telescope/SHOC Johnson B light curve data, together with the GTC/OSIRIS spectra, all folded on the orbital period. A systematic search of the parameter space was performed on the inclination, orbital separation, centre-of-mass redshift, temperatures and surface potential of both stars until $\chi^2$ was globally minimised. Uncertainties in the parameters represent standard 1-$\sigma$ formal errors. See the Methods section for details on the modelling process.}
\end{table}

\vfill\eject

\begin{figure*}[p]
\center
\resizebox{10cm}{!}{\includegraphics{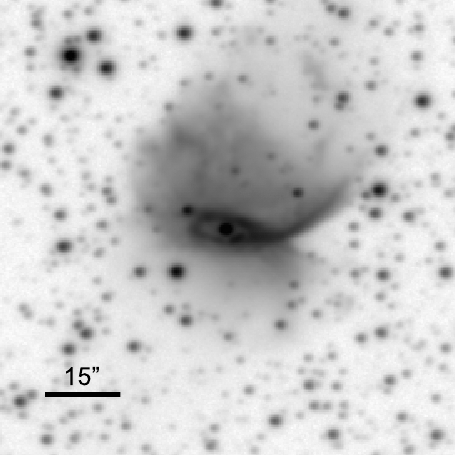}}
\caption{Close-up view of the bipolar planetary nebula Henize~2--428. This 2 hour-deep image in H$\alpha$ 656.3~nm was observed with the INT/WFC. North is up and East is to the left.}
\label{Fp}
\end{figure*}

\begin{figure*}[p]
\center
\resizebox{14cm}{!}{\includegraphics{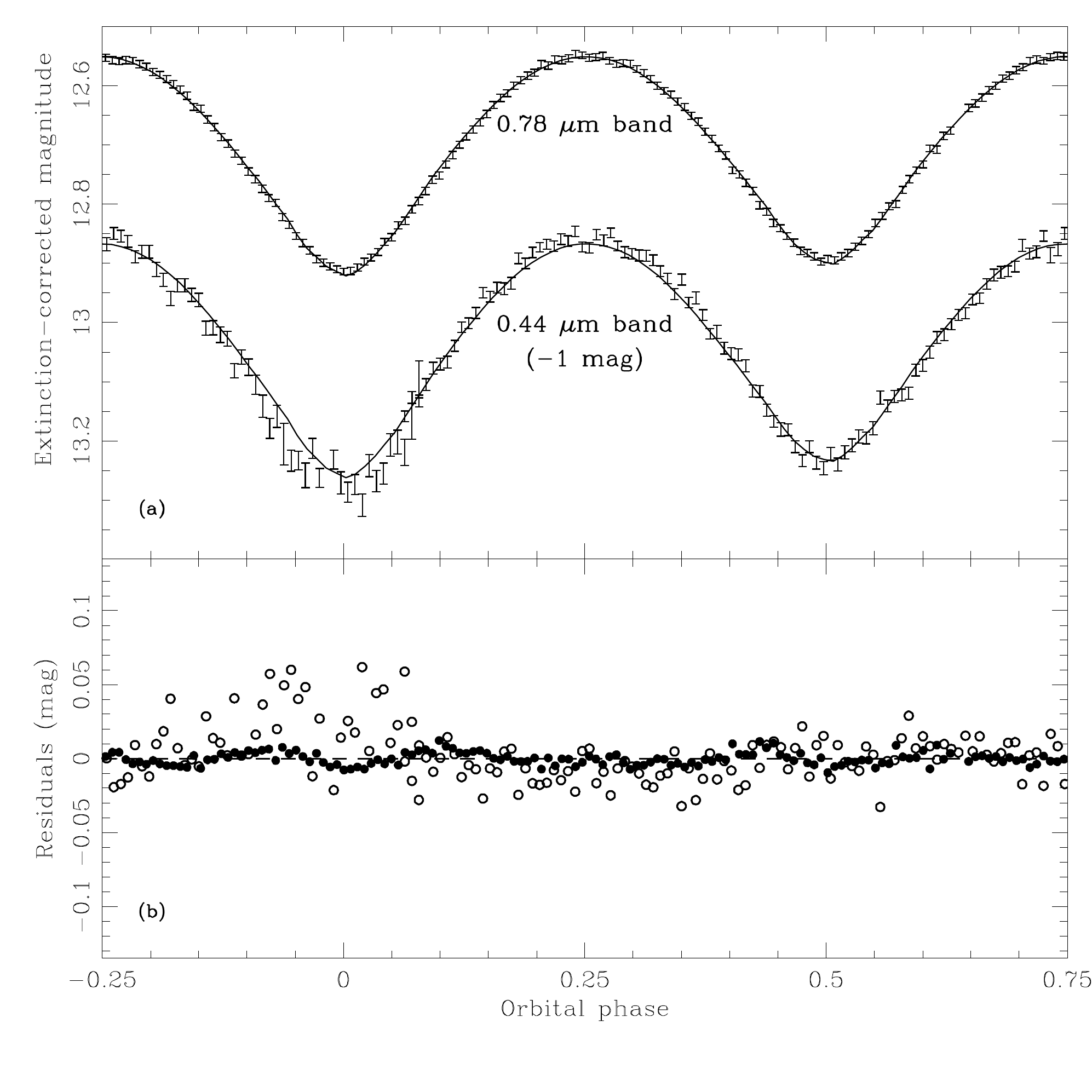}} 
\caption{{\it (a)} Light curve measurements and model. {\it (a)} Light curves of Henize~2--428 in the Johnson B and Sloan i filters (0.44 and 0.78 $\mu$m, respectively) and model, along with {\it (b)} their respective residuals. The B-band data have been shifted up by 1 magnitude for displaying purposes. The data are shown here folded on the orbital period of the system, 0.1758 day, or 4.2 hours, along with the model (solid line). Error bars represent 1-$\sigma$ formal measurement errors.}
\label{F2}
\end{figure*}

\begin{figure*}[p]
\center
\resizebox{14cm}{!}{\includegraphics{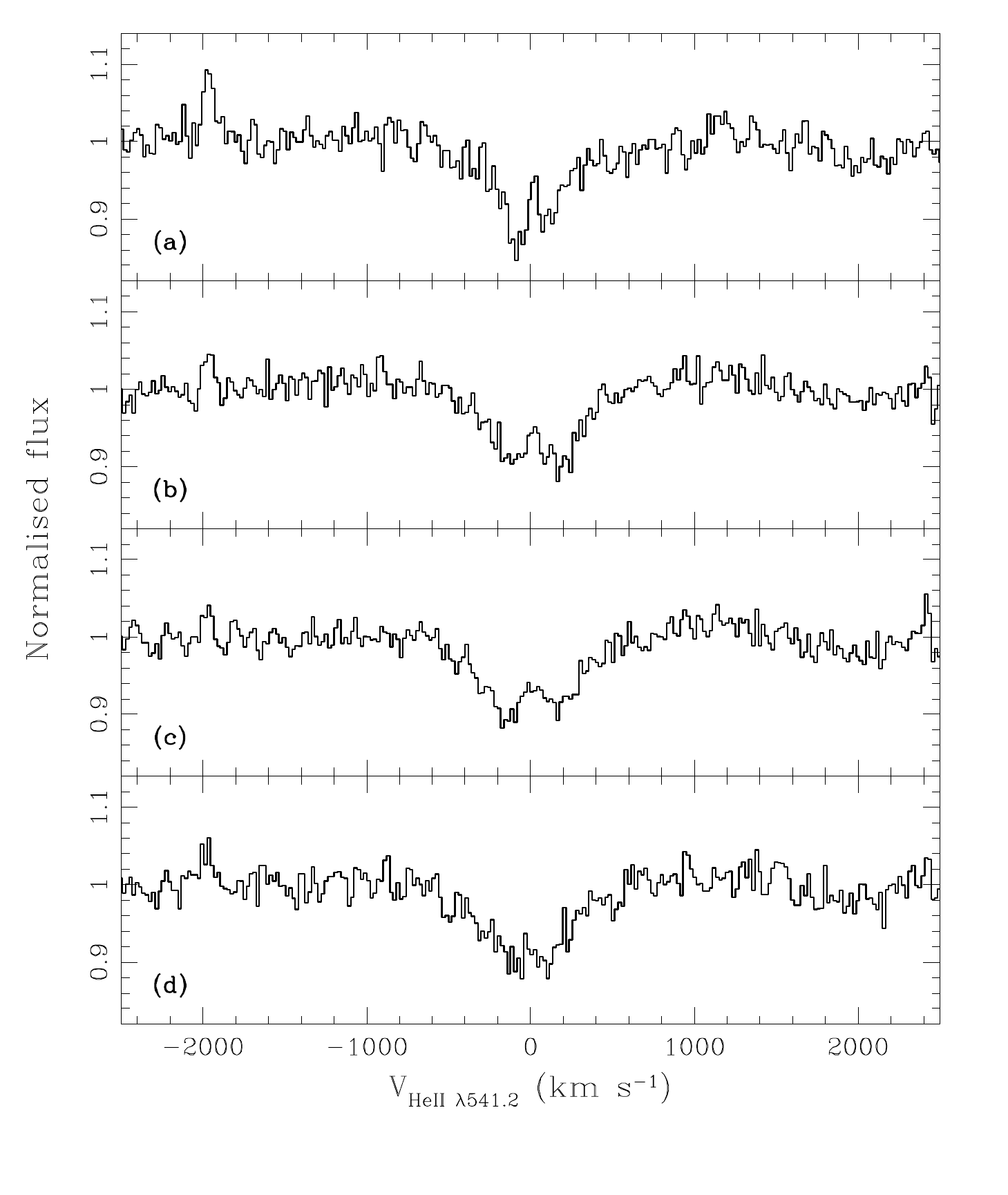}} 
\caption{Time evolution of the spectrum profile of Henize~2-428. The double He {\sc ii} 541.2~nm absorption lines show significant Doppler-shifts in the VLT spectra. The flux is normalized with respect to the continuum. Velocities with respect to the He {\sc ii} 541.2~nm rest wavelength are displayed in the $x$ axis. The top spectrum {\it (a)} corresponds to the night of June 19, 2010, while the three remaining, consecutive spectra were taken on July 8, 2012 and are chronologically ordered from top to bottom {\it (b, c, d)}.}
\label{F3}
\end{figure*}

\begin{figure*}[p]
\center
\resizebox{14cm}{!}{\includegraphics{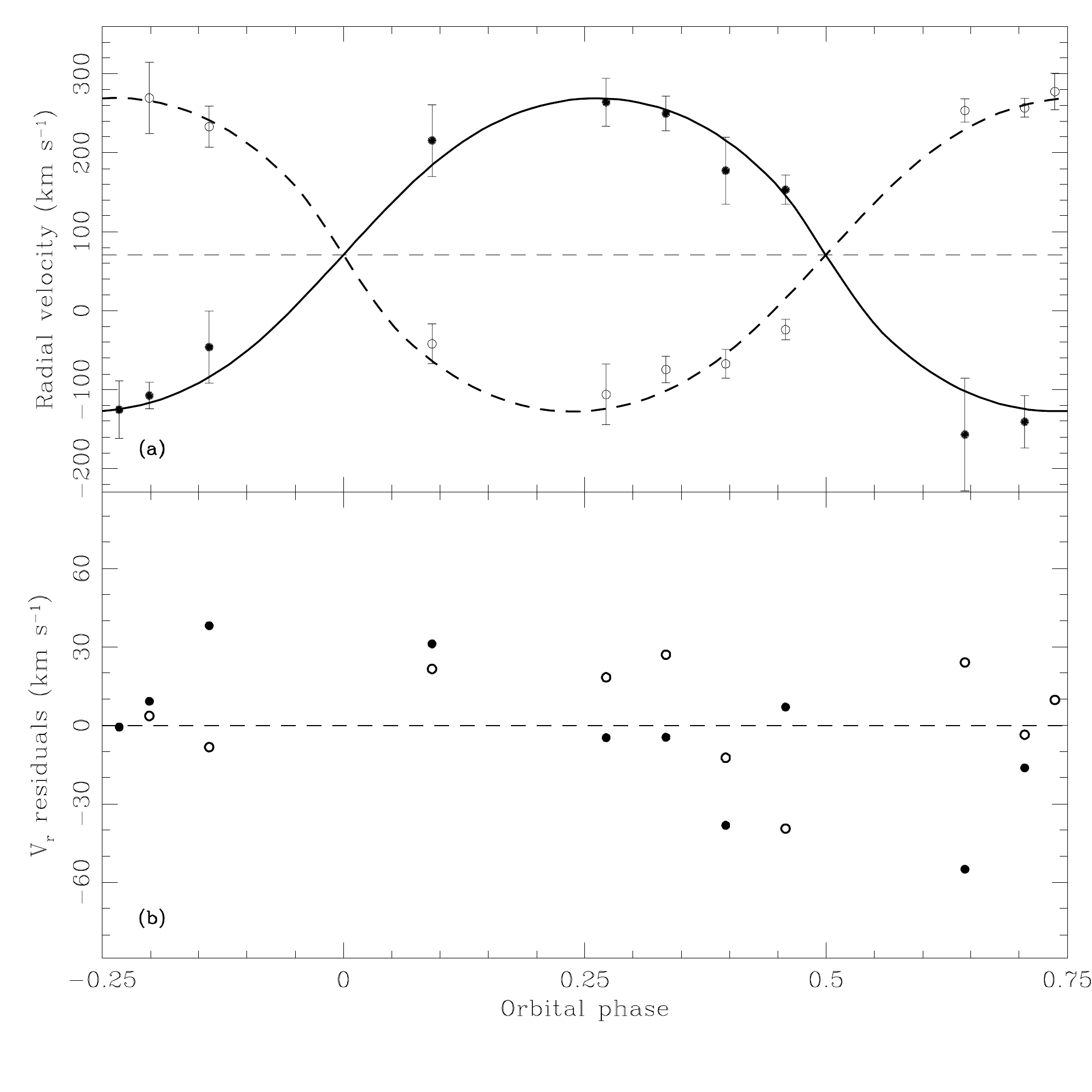}} 
\caption{Radial velocity measurements and orbit solution. {\it (a)} Radial velocity curves of the central stars of Henize~2--428 obtained with GTC/OSIRIS on August 11, 2013, and model, along with {\it (b)} their respective residuals. The data have been folded on the 4.2 hour period determined in the text. The primary star is depicted by black points and the secondary star by white ones, and the dashed horizontal line represents the systemic velocity. Error bars represent 1-$\sigma$ formal measurement errors.}
\label{F4}
\end{figure*}

\clearpage
\newpage

{\center\section*{Methods}}

\subsection{Information on the observational data}

Four short series of I-band time-resolved photometry of Henize~2--428 with MEROPE\citemet{davignon04} on the Mercator telescope\citemet{raskin04} on La Palma on August 28 and 30, 2009. They showed a photometric variability as large as $\sim$0.36~mag between series, so the system was monitored for a single, 4-hour interval on the night of September 2, 2009, and an orbital period was determined. Another similar 4-hour time-series was covered in the Johnson B-band with the SAAO 1m telescope/SHOC on July 11, 2013, and in the Sloan i-band with the INT/WFC on August 2, 2013.  

A single 20~min spectrum with VLT/FORS was secured in the blue range on June 19, 2010, under program ID 085.D-0629(A), and three additional 15 min spectra with the same configuration on the night of July 8, 2012, with program ID 089.D-0453(A). The 1200g grism with a slit width of 0.7 arcsec was used in all cases. The spectra covered the 409-556~nm range. The resulting effective resolution was 0.8 \AA. These 4 spectra taken at different times during an orbit clearly showed radial velocity variations from both stars and established the need for and feasibility of a systematic radial velocity study (see Fig.~3). This was carried out with GTC/OSIRIS on August 11, 2013. The object was monitored for 3.8 hours with a slit width of 0.6 arcsec at parallactic angle. The R2000B grating was used, resulting in a wavelength coverage from 396 to 569.5~nm, and the effective resolution was 1.9 \AA. The spectra were binned once (binning 2$\times$1) in the spatial direction.


\subsection{Data modelling}

The magnitude values form the INT/WFC Sloan i and the SAAO 1m telescope/SHOC Johnson B light curve data were corrected for an extinction of $A_\mathrm{v}$= 2.96$\pm$0.34, recomputed from an available value\citeart{rodriguez01} using a different extinction law\citemet{fitzpatrick04}. We performed a period analysis using Schwarzenberg-Czerny's\citeart{Scwarzenberg96}  analysis-of-variance (AOV) method on the photometric data set with the greatest time coverage (from MERCATOR/MEROPE covering 6.25 hours along three nights, not shown here). The AOV periodogram shows the strongest peak at  $\sim$11.379 cycles d$^{-1}$, which would correspond to a period of 0.0879 day, or 2.1 hours. The orbital period of the system is, however, twice as long,  0.1758$\pm$0.0005 day (4.2 hours), as indicated by the ellipsoidal modulation of the light curve, with the two minima showing similar but different depths.

The ephemerides of the light curves indicated in Table 1 were computed from their respective light curve data. Due to the asymmetry in the radial velocity curves, the corresponding ephemeris of the zero phase was computed from the ephemeris of the Sloan i light curve data, taken 9 nights before, and therefore the associated, accumulated error is much larger.

The orbital and physical parameters of the nucleus of Henize~2--428 were determined by modelling the light curves, together with the radial velocity-curves, all folded on the orbital period, 0.1758 day. The mass ratio $q$ was fixed to 1, as suggested by the amplitude of the radial velocity curves. A systematic search of the parameter space was performed on the inclination, orbital separation, centre-of-mass velocity, temperatures and surface potential of both stars until $\chi^2$ was globally minimised. Given the high effective temperature of both stars, an albedo of 1.0 was used for both components. Gravity brightening and square-root limb darkening coefficients were computed according to the temperature and gravity of each component\citemet{claret11}$^,$\citemet{castelli04}. 

We provide a rough estimate of the distance to Henize~2--428 from the comparison of the model's total luminosity to the dereddened, apparent magnitudes, adopting a bolometric correction according to the stars effective temperatures. The resulting distance is 1.4$\pm$0.4 kpc.

\subsection{B-band data phase determination and alternate model}

Observations in the B-band were taken 22 nights before the data used to determine the orbital phase; error accumulation during this time interval amounts to half an orbital period, thus preventing accurate phase determination of the B-band data. Therefore, actually two possibilities were independently considered and modelled; one where the deepest minimum in both light curves occurs at orbital phase 0 (model shown in the paper), and another one where the deepest minimum is offset by half an orbit.

The results of both models are very similar within uncertainties. These include the inclination, which is confined to a narrow range between 62.2$^\mathrm{o}$ and 63.8$^\mathrm{o}$, still a few degrees away from the $\sim$68$^\mathrm{o}$-inclined equatorial ring of the nebula. The surface potential of both stars are slightly different in this model, and the temperature of the primary is around a thousand K lower than that of the secondary. The total mass is slightly larger but similar to the model shown in the paper, 1.84 M$_\odot$, still above the Chandrasekhar limit even when considering uncertainties. The total luminosity of the system is somewhat larger, 1200 L$_\odot$.

There is not enough solid ground to adopt one model over the other.

%


\bibliographymet{msantander}

\end{document}